\documentclass[10pt,letterpaper,aps,pra,twocolumn,superscriptaddress]{revtex4-1}
\usepackage[latin1]{inputenc}
\usepackage{amsmath}
\usepackage{amsfonts}
\usepackage{amssymb}
\usepackage{bm}
\usepackage[pdftex]{graphicx}
\usepackage[colorinlistoftodos]{todonotes}

\begin{document}
\title{Can a Dove prism change the past of a single photon?}
\author{Miguel A. Alonso}
\affiliation{The Institute of Optics, University of Rochester, Rochester, New York 14627, USA}
\affiliation{Center for Coherence and Quantum Optics, University of Rochester, Rochester, New York 14627, USA}
\author{Andrew N. Jordan}
\affiliation{Department of Physics and Astronomy, University of Rochester, Rochester, New York 14627, USA}
\affiliation{Institute for Quantum Studies, Chapman University, 1 University Drive, Orange, CA 92866, USA}
\affiliation{Center for Coherence and Quantum Optics, University of Rochester, Rochester, New York 14627, USA}
\date{January 6, 2015}
\begin{abstract}
We reexamine the thought experiment and real experiment of Vaidman {\it et al.} \cite{Vaidman1,Vaidman2}, by placing Dove prisms in the nested Mach-Zehnder interferometer arms.  In those previous works, the criterion of whether a single photon was present, or not, was the presence of a ``weak trace", indicating the presence of a nonzero weak value.  This was verified by slightly varying the mirror angle at a given frequency, which was then detected on a position sensitive detector at the oscillation frequency.  We show the presence of the Dove prisms gives identical weak values everywhere to the previous configuration because the prisms change neither the path difference, nor the mode profile in the aligned case. Nevertheless, the same slight variations of the interferometer mirrors now give a signal at the first mirror of the nested interferometer.  We can interpret this result as a misaligned optical interferometer, whose detailed response depends on the stability of the elements, or as the detector coupling to a nonzero effective weak value.
\end{abstract}

\newcommand{\op}[1]{\hat{\bm #1}}                
\newcommand{\ket}[1]{\lvert#1\rangle}
\newcommand{\bra}[1]{\langle#1\rvert}
\newcommand{\pr}[1]{\ket{#1}\bra{#1}}
\newcommand{\ipr}[2]{\langle #1 | #2 \rangle}
\newcommand{\mean}[1]{\left\langle #1 \right\rangle}
\newcommand{\cw}{\circlearrowright}
\newcommand{\ccw}{\circlearrowleft}
\newcommand{\be}{\begin{equation}}
\newcommand{\ee}{\end{equation}}
\newcommand{\bea}{\begin{eqnarray}}
\newcommand{\eea}{\end{eqnarray}}
\newcommand{\ra}{\rangle}
\newcommand{\la}{\langle}

\maketitle
Recent papers by Lev Vaidman and collaborators have explored the possibility of inferring where a photon was in the past, based on a present detector outcome \cite{Vaidman1,Vaidman2}.
This continues the tradition of John Wheeler who proposed and carried out experiments to infer a photon's past trajectory, concluding from a detection event that it came by both paths or by a single path in a Mach-Zehnder interferometer, depending on whether a second beam splitter is placed in the interferometer or not.  This decision can be made {\it after} the photon passes the first beam-splitter, leading to the photon's ``delayed choice" of behaving like a particle or a wave \cite{Wheeler}.  Related interesting results, such as the ``interaction free measurement" (the ability to detect an object's presence with a single photon by detecting an event that would have been impossible had the object been there 
\cite{bomb}),  involve ``counterfactual" reasoning:  If an event could have happened in a given situation, but did not (or its reverse), what can we infer about it?

Vaidman and collaborators have analyzed a Mach-Zehnder interferometer within a Mach-Zehnder interferometer.  They argue that conditioned on certain detector clicks, one can infer the photon's past.  This question does not fall within usual quantum mechanics, and different interpretations give different answers:  Bohr says the question is not well posed - so don't ask it \cite{Bohr}; Wheeler says the photons can retro-actively change their past reality \cite{Wheeler};  Bohm gives a definite trajectory \cite{Bohm}, but it can be ``surrealistic'' \cite{Noam}, etc.

Vaidman has analyzed this situation within the Aharonov-Bergmann-Lebowitz two-state vector formalism \cite{2time}, where there is a forward evolving wavefunction from the prepared single photon state, and a backward evolving wavefunction from the detection event.  This analysis indicates something striking:  the regions of overlap between these forward and backward evolving wave-functions are nonzero in two regions - one is the outer arm of the interferometer, and the other is inside the inner interferometer (see Fig. 1).  This result is interpreted as the photon's past (conditioned on both the pre- and post-selection) being in both the outer arm of the interferometer, as well as the inner arm of the interferometer - despite there being no connection between them!    This interpretation has similarities to the transactional interpretation \cite{TI}.

In order to test this idea, Vaidman has provided a criterion as to how to decide this question:  the photon's past is determined by it leaving a ``weak trace" behind.  This can be checked experimentally by weakly measuring a projection operator on that arm of the interferometer to check the photon's presence or absence.  This criterion also relies on ascribing reality to the weak value - the presence of a non-zero weak value is interpreted as evidence that the photon was present.  This is done in experiment \cite{Vaidman1}, by slightly tilting in an oscillating fashion every mirror in the interferometer with a different frequency.  By detecting the light on a split detector, the signal is then Fourier analyzed, and a peak at that frequency is defined as a weak trace.  Figs.~1(a,b) show the setup used by Vaidman and colleagues, and illustrate the effect of slightly tilting some of the mirrors.
\begin{figure}
  \includegraphics[width=0.4\textwidth]{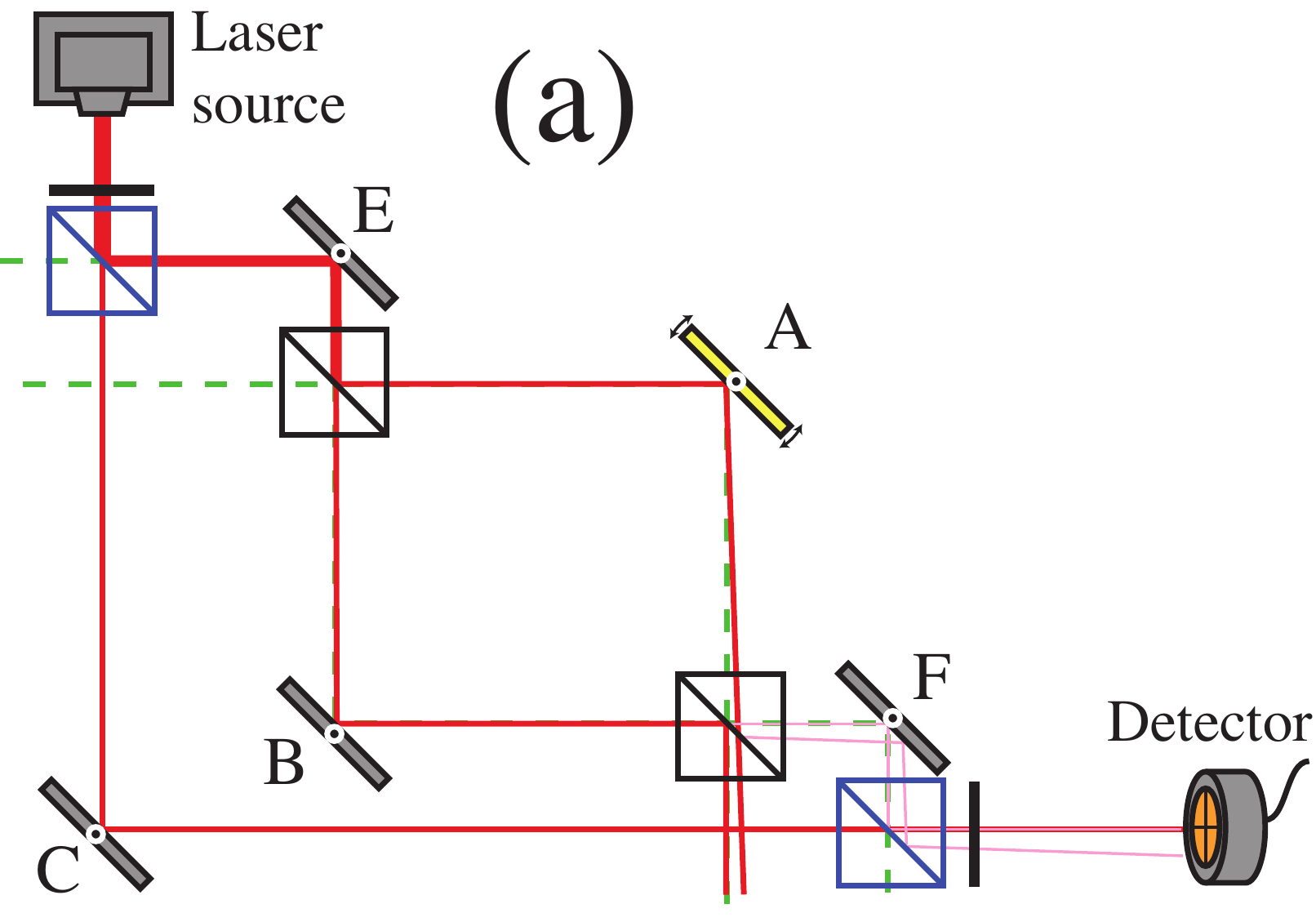}
  \includegraphics[width=0.4\textwidth]{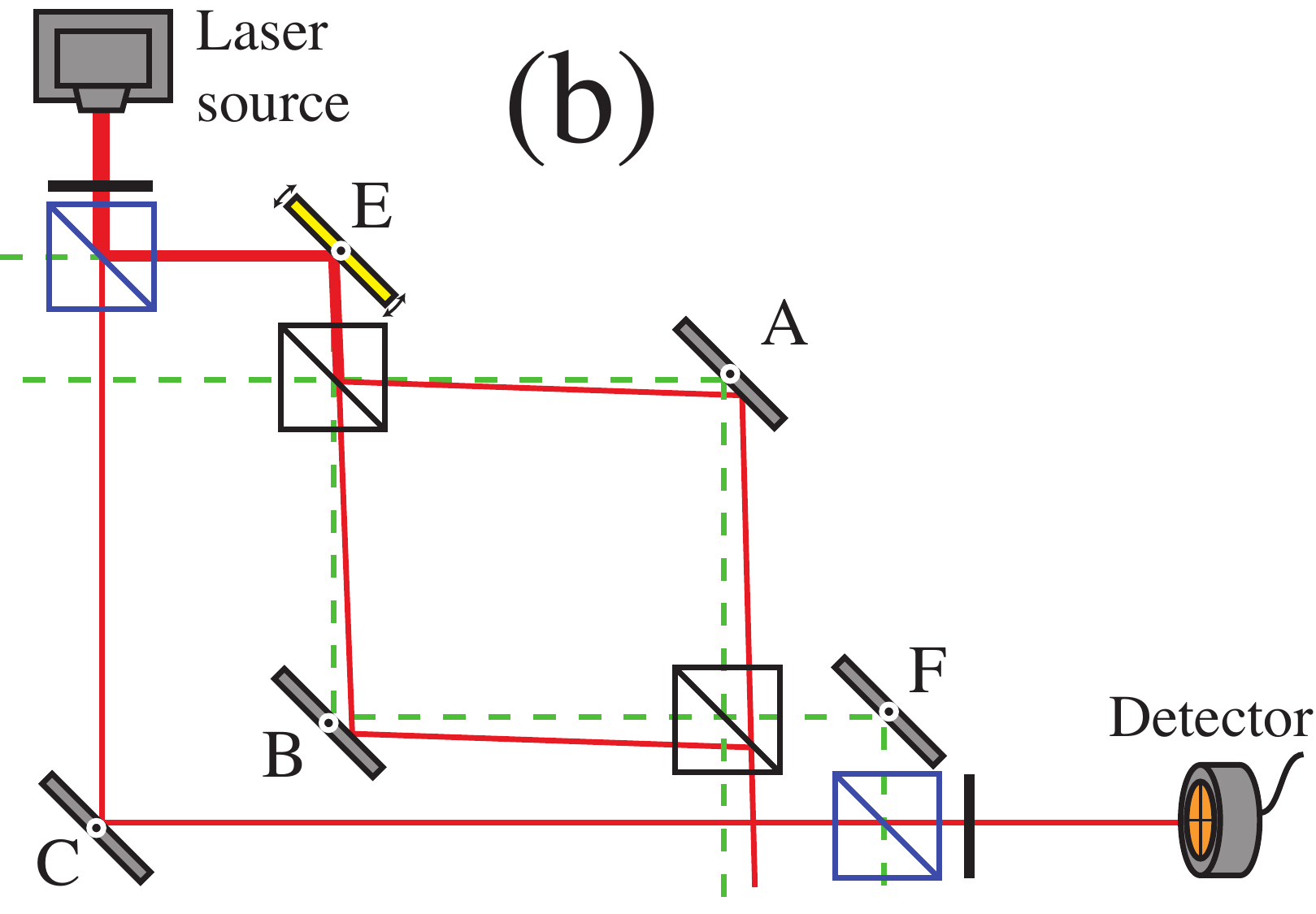}
  \includegraphics[width=0.4\textwidth]{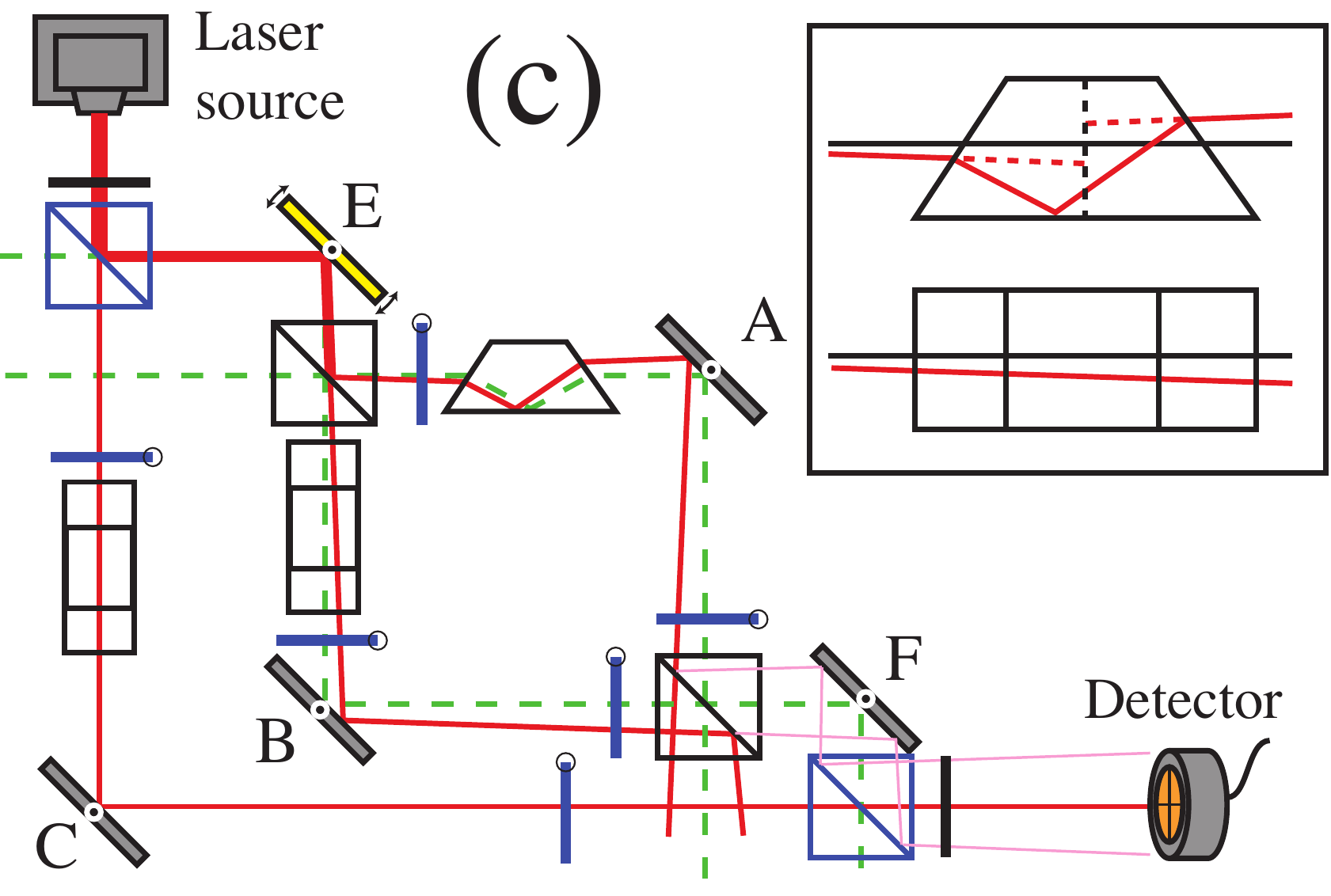}
  \caption{Optical setup used by Vaidman {\it et al.} \cite{Vaidman1,Vaidman2} (a,b) and its modification (c) with Dove prisms inserted on each path. The beam splitters shown in black are 1:1 and polarization-insensitive, while those in blue (top-left and bottom-right) are polarizing ones and act as 2:1 through a suitable choice of the input and output polarizations, selected by polarizers (thick black lines). (a) The red line illustrates the misalignment introduced by tilting mirror A. (b) When mirror E is tilted instead, the two legs going through mirrors A and B bend, but interfere without relative misalignment and cancel exactly before reaching mirror F. (c) When Dove prisms with different orientations are added to all legs, the robustness of the inner Mach-Zehnder interferometer to tilts of mirror E is broken. The blue lines indicate half-wave plates whose fast axes are at $45^{\circ}$ with respect to the interferometer plane (coming out of the plane on the side indicated by a circle), introduced for polarization correction. The inset in (c) shows that a Dove prism acts as a parity operator in one transverse direction but not in the other.}
  \label{fig:state-estimation1}
\end{figure}


There have been several comments concerning Ref.~\cite{Vaidman1}. Refs.~\cite{alignment} points out that alignment is critical: for example, if the mirrors A and B are tilted together, their effect will cancel out, leaving no trace of their (combined) tilts on the detector. A related comments by Svensson criticises associating a (postselected) weak value in a misaligned interferometer to the particle position in an aligned one \cite{comments}.  These interpretational questions have led to a dispute with Zubairy and colleagues, about the actual counterfactuality of a counterfactual communication proposal.   In that dispute, Zubairy claims there can be a secure communication channel between two parties by having both parties select on events that can be interpreted as having no photon inside the communication channel - this also relies on making a claim about where the photon was in the past, given certain detection results \cite{Suhail2}.  Vaidman criticized that proposal because it is a more complicated realization of the nested interferometer \cite{Vaidmanreply}.  Just as he claimed the photon was in the nested interferometer in the past, despite there being no way to enter it, so too, in the Zubairy proposal, he claims that photons are actually in the public communication channel, and the communication is therefore not counterfactual.  This resulted in further debate \cite{Suhail1,Vaidmancomment}. 

The purpose of this paper is to further explore the concept of Vaidman's ``weak trace'' criteria in the modified set-up of Fig.~1(c).  We show the experimental procedure of varying the mirrors to reveal the weak trace is not as simple as it appears.  By adding Dove prisms to the inner interferometer, the weak values remain exactly the same, but the process of wiggling the interferometer mirrors reveals a new situation.  There will now be a weak trace from mirror E, despite the fact the mirror E's weak value is 0.  This fact motivates the title of the paper: If we follow the principle of the weak trace, we must conclude that the photon's presence or absence at mirror E depends on the presence or absence of the (later) Dove prisms.  

{\it Setup.---} We begin our analysis with a review of the two-state vector formalism.  The system is prepared in a state $| \Psi \rangle$ and post-selected in state $\langle \Phi |$.
The selected state is indicated by a detector click in the simplest realization.
The pre-selected state is propagated forward in time, and the post-selected state is propagated backward in time to meet at a point in between - for us, this is in the past.  The weak value \cite{AAV,me} $A_w$ of any operator ${\bf A}$ at the intermediate point may be formally calculated to give,
\be
A_w = \frac{\langle \Phi | {\bf A} | \Psi \rangle}{\langle \Phi | \Psi \rangle}. \label{wvdef}
\ee
We now apply this formula to projection operators $\Pi_j$, where $j$ indicates the various points inside the interferometer.  Using our conventions for the phase shifts acquired by the optical elements, the two-state vector is
\be
\la \Phi | \ \  | \Psi \ra =\frac{1}{\sqrt{3}} (\la {\rm A}| + i \la {\rm B}| - \la {\rm C} |) \ \frac{1}{\sqrt{3}}  (|{\rm A} \ra + i |{\rm B}\ra - |{\rm C} \ra),
\label{tsv}
\ee
which give the weak values,
\begin{eqnarray}
\Pi_{{\rm A},w} &= 1,\,\,\,\Pi_{{\rm B},w} = -1,\,\,\,\Pi_{{\rm C},w} = 1,\nonumber\\
& \Pi_{{\rm E},w} = 0,\,\,\,\Pi_{{\rm F},w} = 0.
\label{wvs}
\end{eqnarray}
It is important to stress that these results are calculated for an aligned interferometer.  In particular, all discussion of the transverse mode structure is suppressed because it is irrelevant - the weak value is a {\it system quantity only}.  

Adding the Dove prisms in Fig. 1(c) does nothing to the above calculations.  Let us see why: If the interferometer has a laser injected into it, prepared in a mode that is symmetric about the optical axis, the beam will be refracted when it enters the prism, bounce off the lower surface, and refract again when it leaves the prism.  The resulting exiting mode is essentially identical to the one that entered, leaving no effect on the light beam.  Consequently, propagating the system state either forward or backward through the Dove prism will be equivalent to free space propagation over a similar optical path length. 

Suppose now that the beam is misaligned with the optical axis. Such misalignment can be introduced by slightly tilting one of the interferometer mirrors.  The paths resulting from such misalignments are represented as red lines in Fig.~1.  Let the transverse coordinates with respect to each aligned, unfolded path, be written as $x$ (within the plane of the interferometer) and $y$ (out of this plane). As can be seen in Fig.~1(c) and its inset, the effect of a Dove prism along the path that includes mirror A is to reflect the mode in $x$ around the optical axis, as well as to reverse the transverse momentum in that direction, $k_x$, while leaving $y$ and $k_y$ unchanged.  Notice that the longer path length and extra internal reflection introduced by this prism will give an extra phase to the photon.  We compensate for this phase by putting other Dove prisms 
in the other two arms of the interferometer to rebalance the path lengths, but we orient these at right angles to the first, so that they act as parity operators in the $y$ direction ({\it i.e.}, out of the interferometer plane).  This way, the paths resulting from mirror deflections in the interferometer plane will be essentially unchanged by these prisms, whereas out-of-plane deflections would be reflected about the optical axis.
As we will now see, the insertion of the prisms changes the stability of the interferometer to mirror tilts.


{\it Results.---} The effect of the Dove prisms on the interferometer's sensitivity to tilts in the mirrors can be understood easily by thinking of the three unfolded paths that join the source with the detector: through mirrors E, A, and F (path EAF), through mirrors E, B, and F (path EBF), and through mirror C (path C). 
Tilting mirror $j$ by a small angle $\alpha_j/2$ (defined as positive if clockwise for mirrors A, E, and F, which face down, and counterclockwise for mirrors B and C, which face up) tilts the reflected beam by $\alpha_j$ and hence gives it a momentum kick by an amount $k\sin\alpha_j\approx k\alpha_j$, so that the effect can be modeled by a phase factor $U_j = \exp(ik\alpha_j x)$. 
Free propagation has the effect of converting momentum kicks into spatial translations due to beam walk-off. 
The three field contributions at the detector plane, due to each of the three paths, are
\begin{eqnarray}
\varphi_{\rm EAF}(x)&\approx&\frac1{\sqrt{3}}\varphi(x-z_{\rm E}\alpha_{\rm E}-z_{\rm A}\alpha_{\rm A}-z_{\rm F}\alpha_{\rm F})\nonumber\\&\times&\exp[ik(\alpha_{\rm E}+\alpha_{\rm A}+\alpha_{\rm F}) x],
\label{EAF} \\
\varphi_{\rm EBF}(x)&\approx& - \frac{1}{\sqrt{3}}\varphi(x-z_{\rm E}\alpha_{\rm E}-z_{\rm B}\alpha_{\rm B}-z_{\rm F}\alpha_{\rm F})\nonumber\\&\times&\exp[ik(\alpha_{\rm E}+\alpha_{\rm B}+\alpha_{\rm F}) x],
\label{EBF} \\
\varphi_{\rm C}(x)&\approx&\frac1{\sqrt{3}}\varphi(x-z_{\rm C}\alpha_{\rm C})\exp(ik\alpha_{\rm C} x), \label{C}
\end{eqnarray}
where $\varphi(x)$ is the field profile for a beam traveling in free space over the same optical distance, and $z_j$ is the optical distance between mirror $j$ and the detector. Note that we neglect phase factors common to all three paths, as well as the dependence in $y$. When the Dove prism is inserted, though, the first of these contribution suffers a change in the sign of $\alpha_{\rm E}$ given the parity flip:
\begin{eqnarray}
\varphi_{\rm EAF}^{({\rm Dove})}(x)&\approx&\frac1{\sqrt{3}}\varphi(x+z_{\rm E}\alpha_{\rm E}-z_{\rm A}\alpha_{\rm A}-z_{\rm F}\alpha_{\rm F})\nonumber\\&\times&\exp[ik(-\alpha_{\rm E}+\alpha_{\rm A}+\alpha_{\rm F}) x].
\label{Dove}
\end{eqnarray}

Note that polarization is not considered in the previous analysis, despite its important effect on the phase shifts each element imposes. Recall that, as in the original experiment \cite{Vaidman1}, the initial and final beam splitters are polarizing ones, and that suitably-oriented polarizers (shown as black lines in Fig.~1) are used at the entrance and exit of the system to guarantee equal weighting of the three paths. The effect of polarization can be addressed by inserting six half-wave plates (indicated as blue lines in Fig.~1(c)), two along each leg, to guarantee that all equivalent optical elements see the same polarization and therefore all legs accummulate the same phase and interfere appropriately. This way, the previous analysis remains valid.

We now calculate the effect of slight tilts in the mirrors on the signal at the detector plane. 
Without the Dove prisms, the state at the detector is a sum of (\ref{EAF}), (\ref{EBF}), and (\ref{C}).
We assume only a symmetric initial meter mode. 
The expected shift of the photon position (the centroid of the beam) when the Dove prisms are not inserted is, in the small angle approximation, independent of tilts of mirrors E and F:
\be
\langle x\rangle=z_{\rm A}\alpha_{\rm A}-z_{\rm B}\alpha_{\rm B}+z_{\rm C}\alpha_{\rm C}.
\ee
When the Dove prisms are inserted, the contribution in (\ref{EAF}) is replaced with that in (\ref{Dove}), so that the centroid now depends on tilts on E, but still not on F:
\be
\langle x\rangle^{({\rm Dove})}=-2z_{\rm E}\alpha_{\rm E}+z_{\rm A}\alpha_{\rm A}-z_{\rm B}\alpha_{\rm B}+z_{\rm C}\alpha_{\rm C}.
\label{Dove-result}
\ee
That is, the stronger dependence is now on $\alpha_{\rm E}$, not only because of the largest numerical factor but because $z_{\rm E}$ is larger than the distances from any other mirror to the detector. Note that the asymmetry between E and F is not due to the fact that the Dove prisms were inserted before mirrors A, B, and C; if they were to be placed right after these mirrors, the previous result would hold with only a change in the sign of $\alpha_{\rm A}$. The signals detected by the split detector are proportional to the above centroids.   We observe that the meter shift at the detector from mirror E is controlled not by the standard weak value $\Pi_{\rm E}$ but by an {\it effective} weak value,
\be
{\tilde \Pi}_{{\rm E},w}  = -2,
\ee
originating from the reflection of the meter profile about the optical axis along path EAF.

{\it Discussion.---}
When a mirror within the system is slightly tilted, it introduces a factor of $\exp(i k\alpha_j x)\approx1+i k\alpha_j x$ to the spatial profile (the meter vector), assumed to be initially an even function ({\it e.g.}, a Gaussian mode). The operator $x$ within the second term of this factor produces a different, odd, spatial transverse mode ({\it e.g.}, a HG01 mode) orthogonal to the first. This odd mode is needed at the split detector to produce a signal. If this odd mode were to follow the same path as the even one from the tilted mirror to the detector, the signal would be proportional to the weak value of the projection operator for the corresponding mirror. This is the case for all mirrors if the Dove prisms are not in place, due to the alignment properties of this system. When the Dove prisms are inserted, though, the even and odd modes traveling from mirror E follow very different paths, since the Dove prism's action as a parity operator produces a sign change for the odd mode only within one of the legs of the inner Mach-Zehnder interferometer. Therefore, the odd mode generated by tilting mirror E is directed in its entirety to the detector, while the even mode leaving this mirror does not reach the detector. The resulting ``weak trace'' is then not proportional to the weak value for the corresponding mirror. Notice that one could also design other systems where the converse is true: a given mirror has a nonzero weak value, but tilting it would not produce a detectable signal on a split detector. One such system would result, for example, from shifting to the left the final beam splitter in the current system, so that it captures instead the other output port of the inner interferometer.

{\it Conclusions.---} We have shown that by adding the Dove prisms to the nested interferometer geometry of Vaidman {\it et al.}, we change the stability of the system.  This results in a misalignment of mirror E now having a detectable effect, whereas previously it did not appear.   The weak values of the various projection operators are identical to those used by those authors in the presence of the Dove prisms, because in the aligned case, these prisms act as identity operators, other than adding optical path length.  The perfect phase balance of the inteferometer is not affected at all.  We showed that the presence of the parity operation indicates that any misalignment of mirror E will leave a weak trace.

What, then, do we conclude from our results?  On one hand, following the arguments of Vaidman, we can say that the presence of the Dove prisms causes the photon to be present (in the past) on mirror E because it now leaves a weak trace - altering its past, depending on the presence of this optical element or not.  Indeed, more radical things were said by Wheeler, who said the presence or absence of a beam splitter retroactively changed the photon path to one or both paths.  Following this line of reasoning, we could further make a ``delayed choice'' experiment, by choosing to either insert the prisms or not {\it after} the photon has passed mirror E - and cause the photon to either leave a trace or not (``it was there or it was not'') on the split detector!  
However, the comparison with the 3 box problem \cite{3box}, where one infers that (-1) particle is in box B because its weak value is -1, shows a danger with this interpretation:  Does one conclude that the Dove prisms cause (-2) particles to be present on mirror E?  Or that if we put the Doves before or after mirror A, that we can change it from +1 particle to -1 particle?
On the other hand, we can simply interpret this result in terms of the stability of the interferometer.  Adding the Dove prisms allows us to break the interferometer stability in the presence of a tilt on mirror E.  

We stress that because the weak values themselves are unchanged, it is possible to (at the same time) make a seperate measurement of the weak value of E using other methods, i.e. a quantum nondemolition measurement with a separate meter, which will then couple to the (zero) weak value.  
So, one conclusion is that the question of leaving a trace on a detector is not as simple as calculating a weak value. The misalignment of an interferometer can bring in other effects that must be accounted for.

How, then, does this connect to the Vaidman/Zubairy debate?  In order to make any claim about the past of the quantum particle, one must have a principle to invoke.  Vaidman's criterion of the weak trace indicating that the photon was at mirrors A and B, but not E and F, has been shown to be sensitive to the details of the interferometer stability.  In response to that, one could adopt the weak values themselves as a new criterion - although to measure them in practice means to slightly break the interference \cite{Suhail1}.
If one adopts instead the criterion of perfect destructive interference as Zubairy does, then it is indeed true that if the interferometer is {\it perfectly} aligned, the sum of paths EAF and EBF cancel exactly not only at the detector, but at all points past the second nested beamsplitter.
Nonetheless, the fact that a slight deviation from perfect alignment in the inner interferometer interferes with path C, making the a photon in the public channel appear without destructive interference, shows the practical fragility of the communication proposal.

{\it Acknowledgements.---}  We give our thanks for lengthy and animated discussions with Lev Vaidman and M. Suhail Zubairy, as well as for their comments on the manuscript.  We both especially wish to thank Jos\'e Javier S\'anchez Mondrag\'on, for his lavish hospitality at the LAOP workshop and conference in Canc\'un, Mexico, where this work was begun. ANJ thanks R. Kastner for discussions.  MAA acknowledges funding from the National Science Foundation under grant PHY-1068325. ANJ acknowledges support from the US Army Research office Grant No. W911NF-09-0-01417.

\end{document}